\documentclass[12pt]{article}
 \newcommand{\be}[1]{\begin{equation}\label{#1}}
 \newcommand{\ba}[1]{\begin{eqnarray}\label{#1}}

 \newcommand{\rd}{{\rm d}}
 
 \newcommand{\re}{{\rm e}}
 \newcommand{\pa}[1]{\left(#1\right)}
 \newcommand{\paq}[1]{\left[#1\right]}
 \newcommand{\pag}[1]{\left\{#1\right\}}
 \newcommand{\av}[1]{\langle#1\rangle}
 \newcommand{\M}{{\rm M_{\rm P}}}
  \newcommand{\z}{{\zeta}}
 \def\ee{\end{equation}}
 \def\ea{\end{eqnarray}}

\usepackage{graphicx,latexsym}
\usepackage{graphicx,epsf}
\usepackage{color}
\usepackage{epsfig}
\usepackage{amsmath}
\usepackage[T1]{fontenc}
\usepackage[utf8]{inputenc}
\usepackage{tikz}
\usepackage{pgfplots}
\usepackage{authblk}
\begin{document}
%%%%%%%%%%%%%
\title{Quantum Cosmology and the Inflationary Spectra from a Non-Minimally Coupled Inflaton}
\author[1,2]{Alexander Y. Kamenshchik\thanks{Alexander.Kamenshchik@bo.infn.it}}
\author[1]{Alessandro Tronconi\thanks{Alessandro.Tronconi@bo.infn.it}}
\author[1]{Giovanni Venturi\thanks{Giovanni.Venturi@bo.infn.it}}
\affil[1]{Dipartimento di Fisica e Astronomia and INFN, Via Irnerio 46,40126 Bologna,
Italy}
\affil[2]{L.D. Landau Institute for Theoretical Physics of the Russian
Academy of Sciences, Kosygin street 2, 119334 Moscow, Russia}

\maketitle

\begin{abstract}
We calculate the quantum gravitational corrections to the Mukhanov-Sasaki equation obtained by the canonical quantization of the inflaton-gravity system. Our approach, which is based on the Born-Oppenheimer decomposition of the resulting Wheeler-DeWitt equation, was previously applied to a minimally coupled inflaton. In this article we examine the case of a non minimally coupled inflaton and, in particular, the induced gravity case is also discussed. Finally, the equation governing the quantum evolution of the inflationary perturbations is derived on a de Sitter background. Moreover the problem of the introduction of time is addressed and a generalized method, with respect to that used for the minimal coupling case, is illustrated. Such a generalized method can be applied to the universe wave function when, through the Born-Oppenheimer factorization, we decompose it into a part which contains the minisuperspace degrees of freedom and another which describes the perturbations.
\end{abstract}

%%%%%%%%%%%%%%%%%%%%%
\section{Introduction}
Inflation \cite{inflation} was originally introduced to overcome the fine tuning problems affecting the old hot big bang cosmology. Today, 40 years after its introduction, by inflation we mean a very articulated framework potentially capable of connecting many aspects of the very early Universe to the present day, low energy, observations \cite{Planck}. Since the microphysics behind inflation is still unknown, people generically speak of the inflationary paradigm and its theoretical description has been declined in many different ways. 

Any inflationary model describing the cosmological evolution during the very early stages of our Universe must supply at least 60 e-folds of accelerated expansion and, during such a phase, the quantum fluctuations of the vacuum, which are believed to generate the seed of the large scale structure we observe today, are stretched beyond the causal horizon giving rise to a nearly scale independent spectrum of perturbations \cite{pert}. Any successful model of inflation must provide a dynamical mechanism which, independently of the initial conditions, satisfies to the above requirements.\\
Moreover, depending on the formulation chosen for the inflationary paradigm, other observable outcomes may be generated by the accelerated phase such as primordial gravitational waves and black holes (which today may constitute part of the Dark Matter content of our Universe \cite{BHform}). Due to its high energy origin inflation can provide an answer to many fundamental problems of modern physics such as the origin of the Dark components of the Universe or the description of Quantum Gravity and can fill the huge gap between the physics at Planck scales down to the Standard Model and classical General Relativity (GR).\\
It is an accepted belief that GR is an effective description of gravity at large distances (low energy). At Planck energies the classical description provided by GR must include new effects arising from quantum mechanics and a new, description of the microscopic world at the Planck length could even be possible \cite{AS}. Quantum effects could generate new operators, irrelevant at low energies, such as higher powers of curvature and any theory sector containing a scalar field (such as the Higgs field \cite{higgsinf}) may couple to gravity non-minimally, drive inflation and/or dynamically affect the ``effective'' Newton's constant. The latter case has been investigated many years before inflation was introduced and was originally called Induced Gravity (IG) since the gravitational field equations were a consequence of (were ``induced by'') the quantum behaviour of some scalar field on the curved background \cite{induced}. Later on \cite{cooper}, it was realised that such a model could in principle drive inflation and become GR at low energies (in the presence of a suitable potential).\\
Induced gravity is thus a natural candidate for the description of the very early Universe, including inflation and quantum effects. The addition of non-perturbative quantum gravitational corrections to such a class of models would lead to an even more complete description of inflationary physics close to Planck energies. In this article we shall consider such a possibility as we canonically quantize the minisuperspace (homogeneous) degrees of freedom (DOF) and then study the evolution of the vacuum fluctuations on the homogeneous background. The method employed was already applied to the GR case in a series of articles \cite{BOarts} and is based on the quantization of the Hamiltonian constraint leading to the Wheeler-DeWitt (WdW) equation for the wave function of the universe \cite{DeWitt}. After a decomposition \`a la Born-Oppenheimer (BO) \cite{BO} for the total universe wave function one is led to a quantum equation for the homogeneous degrees of freedom, which includes the back-reaction of the quantum fluctuations, and an equation for the wave function of each mode of the quantum fluctuations which also depends on the minisuperspace variable. In this context we illustrate a general method for the introduction of the classical time in this latter equation, based on the solution of the Hamilton-Jacobi equation for the minisuperspace variables. This method is a generalisation of that used in the GR case and can be applied to a class of solutions of the homogeneous WdW equation which cannot be non trivially decomposed into a gravitational part and a homogeneous inflaton part \cite{IGwdw}.\\
Finally we apply our method to the de Sitter case which, for IG, is given by a quartic potential. The equation governing the evolution of each mode of the vacuum fluctuation is found to be the same as is obtained for GR and de Sitter (with a constant potential for the scalar field). This result is derived by evaluating the non adiabatic effects emerging from the BO decomposition perturbatively and showing that, at least in the de Sitter case, the inflationary spectra are invariant w.r.t. the Jordan to Einstein frame transition even when the quantum gravitational corrections are included. This result is non trivial and is complementary to that obtained in a previous article, \cite{IGwdw}, where such an equivalence was shown only for the homogeneous DOF. \\
The article is organized as follows. In section 2 we review the basic equations and introduce the formalism. In section 3 we first apply the BO decomposition to the inflaton gravity system and we then illustrate how time can be introduced in this context so as to finally derive the MS equation with quantum gravitational corrections included. In section 3 we apply the formalism to the de Sitter case and finally in section 4 we illustrate our conclusions.

%%%%%%%%%%%%%%%%%%%%%%%%%%%%%%%%

\section{Beyond the Minisuperspace approximation}
Let us consider a non minimally coupled scalar field on a curved, spatially flat, spacetime described by the following action
\be{act0}
S=\int d^4 x\sqrt{-g}\paq{-\frac{U}{2}R+\frac{1}{2}\partial_\mu\phi\partial^{\mu}\phi-V(\phi)}
\ee
where $U=\pa{M^2+\xi\phi^2}$. 
The above action can be decomposed into a homogeneous part plus fluctuations around it. In what follows we shall only consider the scalar fluctuations of the metric. They are associated with the scalar field and can be collectively described in terms of a single, Mukhanov-Sasaki (MS) field $v(x,t)$.
The full lagrangian density governing the evolution of the homogeneous variables and perturbations is given by
\be{act} 
\mathcal{L}=-L^3\pa{3U\frac{a\dot a^2}{N}+6\xi\phi\dot \phi\frac{a^2\dot a}{N}-\frac{a^3\dot \phi^2}{2N}+a^3NV}+\sum_k\mathcal{L}_k
\ee
where the dot denotes the derivative w.r.t. a generic time variable associated with the lapse function $N$ and $\mathcal{L}_k$ is the lagrangian of the k-mode of the Mukhanov-Sasaki variable $v_k$ which here describes scalar perturbations\footnote{A formally identical contribution can be added to describe the tensor perturbations.}. Let us note that, on working in a flat 3-space, and considering both homogeneous and inhomogeneous quantities, one must introduce an unspecified length $L$ (see \cite{BOarts} for more details). In what follows we shall set $L=1$. The lagrangian $\mathcal{L}_k$ takes the form 
\be{pertslag}
\mathcal{L}_k=\frac{1}{2}\pa{v'_k-\omega_k^2v_k^2}
\ee
with
\be{freqdef}
\omega_k^2=k^2-\frac{z''}{z}
\ee
when the conformal time ($N=a$) is chosen. The time dependent mass term for the scalar perturbations $z''/z$ in this context is defined in terms of the homogeneous classical DOF as
\be{zdef}
z\equiv\frac{a^2\phi'}{a'}\pa{1+\frac{6\xi^2 \phi^2}{U}}^{1/2}\pa{1+\frac{\xi a\phi \phi'}{a'U}}^{-1}
\ee
and is a function of time. Moreover the MS variable $v_k$, in the uniform curvature gauge, is
\be{vk}
v_k=\frac{z\,a'}{a\,\phi'}\delta\phi_k,
\ee
where $\delta\phi_k$ is the Fourier transform of the inflaton field fluctuations.\\
The definition of the momenta mixes the velocities of the homogeneous DOF in the minisuperspace approximation (see \cite{IGwdw}) and one has
\be{clmomenta}
\pi_a=-6U a'-6\xi\phi\phi'a,\;\pi_\phi=-6\xi \phi a a'+a^2\phi',\;\pi_k=v'_k.
\ee
Correspondingly the velocities are
\be{vel}
\phi'=\frac{U\pi_\phi-\xi a \phi \pi_a}{a^2\pa{U+6\xi^2\phi^2}},\;a'=-\frac{a\pi_a+6\xi\phi\pi_\phi}{6a\pa{U+6\xi^2\phi^2}},\; v'_k=\pi_k.
\ee
The system Hamiltonian is finally
\be{ham}
\mathcal{H}=\frac{\pi_\phi^2}{2a^2}\frac{U}{U+6\xi^2\phi^2}-\frac{\xi\phi \,\pi_a\pi_\phi}{a\pa{U+6\xi^2\phi^2}}-\frac{\pi_a^2}{12\pa{U+6\xi^2\phi^2}}+a^4V+\sum_k\mathcal{H}_k
\ee
with $\mathcal{H}_k=\frac{1}{2}\pa{\pi_k^2+\omega_k^2v_k^2}$. Given the invariance of the system w.r.t. time reparametrisation, the Hamiltonian $\mathcal{H}$ is zero. \\
The canonical quantisation of the matter-gravity system, then leads to the following Wheeler-DeWitt equation, in the coordinate representation, where, for simplicity, we consider a suitable ordering for the kinetic terms:
\ba{wdw0}
&&\pag{\frac{1}{12U}\partial_A^2+\frac{\xi}{U}\partial_A\partial_F-\frac{1}{2\phi^2}\partial_F^2+a^6\pa{1+\frac{6\,\xi^2\phi^2}{U}}V\right.\nonumber\\
&&\left.+a^2\pa{1+\frac{6\,\xi^2\phi^2}{U}}
\sum_k \hat {\mathcal{H}}_k}\Psi(a,\phi,\paq{v_k})=0
\ea
with $A\equiv \ln a$, $F\equiv \ln \phi$. In the limit $\xi\rightarrow 0$ ($U\rightarrow \M^2$) the above equation becomes
\ba{wdwGR}
\pag{\frac{1}{12a^2\M^2}\partial_A^2-\frac{1}{2a^2\phi^2}\partial_F^2+a^4V+
\sum_k \hat {\mathcal{H}}_k}\Psi(a,\phi,\paq{v_k})=0
\ea
which is its correct GR limit. On the other hand in the limit $\M\rightarrow 0$ ($U\rightarrow \xi\phi^2$) the WdW equation (\ref{wdw0}) becomes
\ba{wdwIG}
&&\pag{\frac{1}{12\xi }\partial_A^2+\partial_A\partial_F-\frac{1}{2}\partial_F^2+a^6\pa{1+6\,\xi}\phi^2V\right.\nonumber\\
&&\left.+a^2\phi^2\pa{1+6\,\xi}
\sum_k \hat {\mathcal{H}}_k}\Psi(a,\phi,\paq{v_k})=0
\ea
and its correct IG limit is recovered.\\
Let us now perform the following Born-Oppenheimer (BO) decomposition where the homogeneous DOF are factorised with respect to the wave function of the perturbations:
\be{BOdec}
\Psi(a,\phi,\paq{v_k})=\Psi_0(A,F)\prod_k\chi_k(A,F,v_k).
\ee
Let us note that each mode of the perturbations is described by the corresponding wave function which also depends on the homogeneous DOF.

%%%%%%%%%%%%%%%%%%%%%%%%%%%%%%%%%%%%%%%%%%%
\subsection{BO decomposition}
The BO decomposition was originally applied in atomic physics and consists of factorising the total wave function of atoms and molecules in a part for the ``slow'' DOF (nuclei) and a part for the ``fast'' DOF (electrons), the latter depending on the ``slow'' variables as well. To the leading order in the adiabatic approximation the BO decomposition then leads to a system of coupled Schr\"odinger equations which can be solved analytically. Non adiabatic terms, at the next to leading order, determine non adiabatic transitions between quantum levels, otherwise neglected in the adiabatic approximation. The same BO approach has been successfully applied to the inflaton-gravity system by usually associating to the scale factor the role played by the nucleus in atomic physics and to matter (homogeneous inflaton and perturbations) that of the electrons. The non adiabatic contributions which arise in the decompositions are, in this context, associated with the quantum gravitational effects. Such effects in the common semiclassical treatment of the evolution of inflationary perturbations are neglected.\\ 
In contrast with \cite{BOarts}, where only the scale factor dependence was factorised, here we followed a more general approach which, in principle, can be applied to systems where the wave function of the ``slow'' (gravitational) DOF cannot be, non trivially, factorised. Moreover in scalar-tensor theories the role of the scalar field (besides being the inflaton) is tightly intertwined with gravity since it dynamically determines Newton's constant and ``induces'' its dynamics through quantum effects.\\
In order to proceed with the BO decomposition let us first rewrite the WdW equation in a compact form as
\be{wdw1}
\pag{\sum_{\alpha,\beta=1,2}G^{\alpha\beta}\partial_\alpha\partial_\beta+a_0^6{\rm e}^{6A}{\mathcal V}+a_0^2{\rm e}^{2A}h
\sum_k \hat {\mathcal{H}}_k}\Psi(a,\phi,\paq{v_k})=0
\ee
where $X=(A,F)$ ($X^1=A$, $X^2=F$), $\partial_{\alpha}\equiv \partial_{X^{\alpha}}$ and 
\be{Gdef}
G\equiv\frac{1}{2}
\left(
\begin{array}{cc}
\pa{6\xi}^{-1}&1\\
1&-g
\end{array}
\right)
\ee
is the metric of the homogeneous minisuperspace.
Moreover let us set
\be{defs}
 g=\frac{U}{\xi\phi^2},\;h=\frac{U+6\,\xi^2\phi^2}{\xi},\;{\mathcal V}=h V
\ee
and from here on we shall use the Einstein summation convention in order to keep the notation as compact as possible.\\
The BO decomposition is performed by splitting the total wave function using the ansatz (\ref{BOdec}). Then an equation for the homogeneous wave function $\Psi_0$ can be obtained by projecting out the inhomogeneous DOF i.e. by contracting the WdW equation with $\prod_k \chi_k^*(A,F,v_k)$ and integrating over $\prod_k \rd v_k$. The resulting equation is 
\ba{homeq0}
&&G^{\alpha\beta}\left\{\partial_\alpha\partial_\beta+\sum_k\paq{2\av{\chi_k|\partial_\alpha\chi_k}\pa{\partial_\beta+\sum_{j\neq k}\av{\chi_j|\partial_\beta\chi_j}}+\av{\chi_k|\partial_\alpha\partial_\beta\chi_k}}\right\}\Psi_0\nonumber\\
&&+\pa{a_0^6{\rm e}^{6A}{\mathcal V}+a_0^2{\rm e}^{2A} h\sum_{k}\av{\chi_k|{\hat {\mathcal H}}_k|\chi_k}}\Psi_0=0
\ea
where 
\be{defprod}
\av{\chi_k|\hat O|\chi_k}\equiv \int_{-\infty}^{+\infty}\rd v_k\,\chi_k^*(a,\phi,v_k)\,\mathcal{R}(\hat {O})\, \chi_k(a,\phi,v_k)
\ee
and $\mathcal{R}(\hat {O})$ is the coordinate representation of the operator $\hat O$. Henceforth, in order to keep the notation compact, we shall use the same notation for quantum operators independetly of the representation used. 
This latter equation correctly reproduces that for minisuperspace \cite{IGwdw} when one neglects the back-reaction of the inhomogeneities on the homogeneous part in the above equation (\ref{homeq0}). In the present context, the back-reaction is given by the semiclassical contribution of the energy density of the inhomogeneities and consists in the sum of the averaged hamiltonians $H_k$ plus the non adiabatic contributions which describe the quantum gravitational effects. These contributions are expected to be small during inflation when the homogeneous inflaton energy density is usually assumed to be, by far, the leading contribution.\\
One then finds the equations for the modes $\chi_k$. These equations can be obtained by multiplying the gravitational equation by $\chi_k$ and then subtracting the WdW equation multiplicated by $\prod_{j\neq k}\chi_j^*$ and integrated over $\prod_{j\neq k} \rd v_j$. \\
The resulting equation is 
\ba{keq0}
&&\!\!\!\!\!\!\!\!\!\!G^{\alpha\beta}\left\{\phantom{\frac{A}{B}}\!\!\!\!\!\!2\pa{\partial_\alpha\Psi_0}\pa{\partial_\beta-\av{\chi_k|\partial_\beta \chi_k}}\chi_k+\Psi_0\pa{\partial_\alpha\partial_\beta-\av{\chi_k|\partial_\alpha\partial_\beta\chi_k}}\chi_k\right.\nonumber\\
&&\!\!\!\!\!\!\!\!\!\!\left.+2\Psi_0\pa{\sum_{i\neq k}\av{\chi_i|\partial_\alpha \chi_i}}\pa{\partial_\alpha-\av{\chi_k|\partial_\alpha\chi_k}}\partial_\beta \chi_k
\right\}\nonumber\\
&&\!\!\!\!\!\!\!\!\!\!+a_0^2{\rm e}^{2A}h\Psi_0\pa{\hat {\mathcal H}_k-\av{\chi_k|\hat {\mathcal H}_k|\chi_k}}\chi_k=0.
\ea
Let us now define the recurrent expression $\av{\chi_k|\hat O| \chi_k}\equiv\av{\hat O}_k$. The expression (\ref{keq0}) is the equation for the wave function of the $k$-mode of the MS field and in the present form also contains the dependence on the modes different from $k$. The equations (\ref{homeq0}) and (\ref{keq0}) are equivalent to the WdW equation (\ref{wdw1}). They can be simplified by re-phasing $\Psi_0$ and $\chi_k$ as follows:
\be{rephase}
\Psi_0\equiv \tilde\Psi_0 {\rm e}^{i \theta (A,F)}\,,\chi_k\equiv {\rm e}^{-i \theta_k (A,F)}\tilde\chi_k
\ee
with
\be{phase}
\theta (A,F)\equiv i\sum_j\int^{A,F}\av{\partial_\alpha}_j d\bar X^{\alpha}\,,\theta_k (A,F)\equiv i\int^{A,F}\av{\partial_\alpha}_kd\bar X^{\alpha}.
\ee
Let us note that the above line integrals are independent of the contour of integration chosen provided no singularities are present in the domain of integration (and this is generally the case) \cite{BO}. Moreover one has 
\be{gradph}
\partial_{\alpha}\theta (X)=i \sum_j \av{\partial_\alpha}_j\,,\; \partial_{\alpha}\theta_k (X)=i \av{\partial_\alpha}_k.
\ee
In terms of the re-defined wave functions the homogeneous equation (\ref{homeq0}) takes the form
\be{homeq2}
G^{\alpha\beta}\partial_\alpha\partial_\beta\tilde \Psi_0+\pa{\re^{6A}{\mathcal V}+\re^{2A}h\sum_k\av{\hat {\widetilde{\mathcal H}}_k}_k}\tilde \Psi_0=G^{\alpha\beta}\sum_k\av{\partial_\alpha\tilde\chi_k|\partial_\beta\tilde\chi_k}\tilde \Psi_0
\ee
where $\av{\hat{\widetilde O}}_k\equiv \av{\tilde \chi_k|\hat O|\tilde \chi_k}$ and the r.h.s. contains the quantum gravitational effects on the total back-reaction of inhomogeneities for the homogeneous background. On neglecting such inhomogeneities one recovers the WdW equation for the minisuperspace variables.\\
The equation for the perturbations finally becomes
\be{keq2}
\left\{G^{\alpha\beta}\paq{2\frac{\partial_\alpha\tilde\Psi_0}{\tilde \Psi_0}\partial_\beta+\pa{\partial_\alpha\partial_\beta-\av{\widetilde{\partial_\alpha\partial_\beta}}_k}}+a_0^2{\rm e}^{2A}h(F)\pa{\hat{\mathcal H}_k-\av{\hat{\widetilde{\mathcal H}}_k}_k}\right\}\tilde\chi_k=0
\ee
and, in contrast with (\ref{keq0}), it only contains a single $k$-mode. Therefore, from here on, we shall omit the external subscript $k$ to keep the notation compact ($\av{\hat {\widetilde O}}_k\rightarrow \av{\hat {\widetilde O}}$).\\
Let us note that the expression $G^{\alpha\beta}(2\partial_\alpha\tilde\Psi_0/\tilde \Psi_0)\partial_\beta$ is related to the introduction of time \cite{BOarts}. It is given by 4 contributions:
\be{time0}
2G^{\alpha\beta}\frac{\partial_\alpha\tilde\Psi_0}{\tilde \Psi_0}\partial_\beta=\frac{1}{6\xi}\frac{\partial_A\tilde \Psi_0}{\tilde\Psi_0}\partial_A+\frac{\partial_F\tilde \Psi_0}{\tilde\Psi_0}\partial_A+\frac{\partial_A\tilde \Psi_0}{\tilde\Psi_0}\partial_F-g(F)\frac{\partial_F\tilde \Psi_0}{\tilde\Psi_0}\partial_F.
\ee
We observe that
\be{timea}
\pa{\frac{1}{6\xi}\frac{\partial_A\tilde \Psi_0}{\tilde\Psi_0}+\frac{\partial_F\tilde \Psi_0}{\tilde\Psi_0}}\partial_A=\frac{i}{\tilde\Psi_0}\pa{\frac{a \hat \pi_a+6\xi\phi \hat \pi_\phi}{6\xi}\tilde\Psi_0}\;a\,\partial_a
\ee
and
\be{timephi}
\pa{\frac{\partial_A\tilde \Psi_0}{\tilde\Psi_0}-g(F)\frac{\partial_F\tilde \Psi_0}{\tilde\Psi_0}}\partial_F=\frac{i}{\tilde\Psi_0}\pa{\frac{\xi \phi a\hat \pi_a-U\hat \pi_\phi}{\xi\phi}\tilde \Psi_0}\phi\,\partial_\phi.
\ee
In the semiclassical limit, to the leading order in $\hbar$, quantum operators can be replaced by their classical counterparts leading to
\be{semi0}
\pa{a \hat \pi_a+6\xi\phi \hat \pi_\phi}\tilde\Psi_0\simeq-6 \xi a\,a'h\,\tilde\Psi_0\,,\;\pa{\xi \phi a\hat \pi_a-U\hat \pi_\phi}\tilde\Psi_0\simeq -\xi a^2\phi'h\,\tilde\Psi_0,
\ee
where the quantities on the r.h.s. in (\ref{semi0}) are the classical (time dependent) variables. 
%Moreover one has
%\be{dertime}
%-i\partial_A\tilde\Psi_0=\pa{-6a a'U-6\xi\phi\phi'a^2}\tilde\Psi_0\,,\,-i\partial_F\tilde\Psi_0=\pa{-6\xi\phi^2 a %a'+a^2\phi\phi'}\tilde\Psi_0
%\ee
Therefore, in such a limit,
\be{timea2}
\pa{\frac{1}{6\xi}\frac{\partial_A\tilde \Psi_0}{\tilde\Psi_0}+\frac{\partial_F\tilde \Psi_0}{\tilde\Psi_0}}\partial_A=G^{\alpha1}\frac{\partial_\alpha\tilde \Psi_0}{\tilde\Psi_0}\partial_1\simeq -i\pa{ a^2a'h}\,\partial_a
\ee
\be{timephi2}
\pa{\frac{\partial_A\tilde \Psi_0}{\tilde\Psi_0}-g(F)\frac{\partial_F\tilde \Psi_0}{\tilde\Psi_0}}\partial_F= G^{\alpha2}\frac{\partial_\alpha\tilde \Psi_0}{\tilde\Psi_0}\partial_2\simeq-i\pa{ a^2\phi'h}\partial_\phi.
\ee
and
\be{timelo}
2G^{\alpha\beta}\frac{\partial_\alpha\tilde\Psi_0}{\tilde \Psi_0}\partial_\beta\tilde\chi_k(a,\phi,v_k)\simeq-ia^2h\frac{\partial}{\partial \eta}\tilde\chi_k(a(\eta),\phi(\eta),v_k),
\ee
where the homogeneous variables must be evaluated on the classical trajectory $\tilde\chi_k(a(\eta),\phi(\eta),v_k)\equiv \tilde\chi_k(\eta,v_k)$. 
%The equation for each mode then takes the form
%\be{keq3}
%\left\{-i\frac{\rd}{\rd \eta}+\frac{G^{\alpha\beta}}{a^2h}\pa{\partial_\alpha\partial_\beta-\av{\widetilde{\partial_\alpha\partial_\beta}}_k}+\pa{\hat{\mathcal H}_k-\av{\hat{\mathcal H}_k}}\right\}|\tilde\chi_k(\eta,v_k)\rangle=0.
%\ee
%%%%%%%%%%%%%%%%%%%%%%%
\subsection{Introduction of time}
As we already pointed out the emergence of time in equation (\ref{keq2}) is related to the derivative of the homogeneous wave function and is therefore a consequence of the BO decomposition. We have also observed (see (\ref{timelo})) that the emerging ``flow'' of the time is defined by the trajectories in the $(A,F)$ manifold (i.e. the configuration space of the homogeneous variables) described by the tangent vector 
\be{tantime0}
\partial_\eta=\eta^A\partial_A+\eta^F\partial_F\equiv\eta^{\alpha}\partial_\alpha,
\ee
where $\eta^A$ and $\eta^F$ are functions defined on the configuration space and corresponding to the classical velocities $\eta^A=\frac{\partial A_{\rm cl}}{\partial \eta}$ and $\eta^F=\frac{\partial F_{\rm cl}}{\partial \eta}$. The integral curves $(A(\eta),F(\eta))$, solutions of the system
\be{intcurv}
\left\{
\begin{array}{l}
\frac{\rd A}{\rd \eta}=\eta^A(A,F)\\
\frac{\rd F}{\rd \eta}=\eta^F(A,F)
\end{array}
\right.,
\ee
represent the classical solutions and the corresponding tangent defines the (classical) time flow. The solutions of (\ref{intcurv}) depend on two integration constants. The resulting curves form a congruence on the configuration space (minisuperspace).\\
Let us note that the ``emergence'' of time is associated with some classical limit of the state described by the homogeneous wave function $\Psi_0$. If the matter-gravity system maintains a purely quantum behaviour a classical time cannot be introduced and no real advantage can be obtained from the BO approach. For example, a well defined classical behaviour in minisuperspace is recovered in the leading order of the WKB approximation or in the large $a$ limit for some quantum solutions to the homogeneous WdW equation \cite{IGwdw}. The introduction of time depends on the quantum fluctuations around the classical trajectory due to the intrinsic quantum nature of the matter-gravity system. When such fluctuations are small they can be treated perturbatively and the classical limit is well defined. The presence of large quantum fluctuation destroys the classical evolution and signals that the system is in a highly quantum (non-classical) state. 
%%%%%%%%%%%%%%%%%%%%%%%%%%%%%%%%
\subsection{Hamilton-Jacobi equation}
In order to obtain the classical flow of the time, one needs the functions $\eta^\alpha$, defined over the configuration space (and corresponding to the minisuperspace). These functions can be calculated from the general solution of the classical Hamilton-Jacobi (HJ) equation. From the classical Hamiltonian $\mathcal{H}=\mathcal{H}\pa{\pi_a,\pi_\phi,a,\phi}$, given by expression (\ref{ham}) (without including the inhomogeneities), we derive the following HJ equation for the HJ function $W(a,\phi)$ 
\be{HJeq}
\mathcal{H}\pa{\partial_a W,\partial_\phi W,a,\phi}=0.
\ee
An exact general solution for (\ref{HJeq}) can be obtained in the IG case for potentials of the form $V=\lambda M^{4-n}\phi^n$ starting from the ansatz
\be{ansHJ}
W=\nu\ln\frac{a}{a_0}+\ln \omega(x),
\ee
where $x\equiv a^3\phi^{\frac{n+2}{2}}$. Thus,
\be{derW}
\partial_A W=\nu+3\frac{\rd \ln \omega}{\rd \ln x},\;\partial_F W=\frac{n+2}{2}\frac{\rd \ln \omega}{\rd \ln x}
\ee
and the HJ equation becomes
\ba{HJ2}
&&\pa{\frac{\rd \ln \omega}{\rd \ln x}}^2\paq{\frac{3}{4\xi}-\frac{\pa{n-4}^2}{8\pa{1+6\xi}}}+\pa{\frac{\rd \ln \omega}{\rd \ln x}}\paq{\frac{1}{\xi}+\frac{n-4}{1+6\xi}}\frac{\nu}{2}+\nonumber\\
&&+\paq{\frac{\nu^2}{12\xi\pa{1+6\xi}}-\lambda M^{4-n}x^2}=0.
\ea
This first order differential equation can be solved algebraically for $\rd \ln \omega/\rd \ln x$ and then integrated to obtain
\be{omegasol}
\omega(x)=\widetilde D\, x^{\widetilde A}\exp\paq{\pm\pa{\sqrt{\widetilde B+\widetilde C x^2}-\sqrt{\widetilde B}\tanh^{-1}\sqrt{1+\frac{\widetilde C}{\widetilde B}x^2}}}
\ee
with
\be{Atilde}
\widetilde A=-\frac{\paq{1+\frac{\pa{n-4}\xi}{1+6\xi}}}{\paq{1-\frac{\xi\pa{n-4}^2}{6\pa{1+6\xi}}}}\frac{\nu}{3},
\ee
\be{Btilde}
\widetilde B=\widetilde A^2-\frac{2\nu^2}{3\paq{6\pa{1+6\xi}-\xi\pa{n-4}^2}},
\ee
\be{Ctilde}
\widetilde C=\frac{4\xi\lambda M^{4-n}}{3\paq{1-\frac{\xi\pa{n-4}^2}{6\pa{1+6\xi}}}}
\ee
and $\widetilde D$ is an integration constant. In the $n=4$ limit the expressions above are further simplified and, in particular, one obtains
\be{domega4}
\frac{\rd \ln \omega}{\rd \ln x}=-\frac{\nu}{3}\pm\sqrt{\frac{2\xi\nu^2}{3\pa{1+6\xi}}+\frac{4\xi\lambda x^2}{3}}.
\ee
The classical velocities can be obtained from (\ref{vel}) with $\pi_a=\partial_a W$ and $\pi_\phi=\partial_\phi W$. For the $n=4$ case one has
\be{vel4}
\frac{\phi'}{\phi}=-\frac{\nu}{x^{2/3}\pa{1+6\xi}},\;\frac{a'}{a}=-\frac{1}{6\xi x^{2/3}}\pa{\frac{\nu}{1+6\xi}+3\frac{\rd \ln \omega}{\rd \ln x}}.
\ee
The constant $\nu$ parametrises different sets of trajectories on the configuration space, the de Sitter (inflationary) attractor trajectory corresponding to $\nu=0$. The expressions (\ref{vel4}) then take the form
\be{vel4dS}
\frac{\phi'}{\phi}=0,\;\frac{a'}{a}=\sqrt{\frac{\lambda}{3\xi}}\,a\,\phi.
\ee

%%%%%%%%%%%%%%%%%%%%%%%%%%%%%%%%
\subsection{Auxiliary vector}
We already observed that
\be{tantime}
\partial_\eta\sim\frac{2iG^{\alpha\beta}}{a^2h}\frac{\partial_{\alpha}\tilde \Psi_0}{\tilde \Psi_0}\partial_\beta
\ee
where the approximate equality becomes exact in the semiclassical limit and to the leading order in $\hbar$. Higher order contributions in $i\partial_{\alpha}\tilde \Psi_0/\tilde \Psi_0$ must be interpreted as quantum gravitational effects related to the definition of time and, if small, can be treated perturbatively. To the leading order one has
\be{wkbtime}
\partial_\eta\equiv \frac{2G^{\alpha\beta}}{a^2 h}\pa{\partial_\alpha W}\partial_\beta=\eta^\beta \partial_\beta,
\ee
where $W$ is a solution to the HJ equation, and the so-called WKB time is recovered. 
On the other hand the above definition of the time flow can be applied to more general solutions of the homogeneous equation which substantially differ from the semiclassical ones and then need a more careful treatment.\\
Once time is formally introduced by (\ref{tantime}) one recovers, to the leading order, the Schr\"odinger equation governing the evolution for the wave function of the inflationary perturbations. Such an equation is equivalent to the Mukhanov-Sasaki (MS) equation for the operator $\hat v_k$ calculated on a classical background (indeed it is the same equation but in the Schr\"odinger representation).\\
If one is interested in calculating the quantum gravitational corrections to the semiclassical MS equation things are more involved. From the definition (\ref{tantime0}) of the time flow one can introduce an ``auxiliary'' vector satisfying
\be{orttime}
\partial_\tau\equiv \tau^A\partial_A+\tau^F\partial_F\equiv \tau^{\alpha}\partial_\alpha\;,{\rm with}\quad \paq{\partial_\eta,\partial_\tau}=0
\ee
where the normalization of $\partial_\tau$ is unspecified and can be fixed arbitrarily. Let us note that $\partial_\tau$ is not defined in a unique way and, for example, $\partial_\tau+c\,\partial_\eta$ still satisfies the condition (\ref{orttime}).
The components of the auxiliary vector, by definition, must satisfy the following equations
\be{auxdef}
\left\{
\begin{array}{l}
\eta^A\pa{\partial_A\tau^F}+\eta^F\pa{\partial_F\tau^F}-\tau^A\pa{\partial_A\eta^F}-\tau^F\pa{\partial_F\eta^F}=0\\
\eta^A\pa{\partial_A\tau^A}+\eta^F\pa{\partial_F\tau^A}-\tau^A\pa{\partial_A\eta^A}-\tau^F\pa{\partial_F\eta^A}=0.
\end{array}
\right.
\ee
In the $n=4$ case $\eta^{\alpha}=\eta^{\alpha}(x)$ with $x=(a\phi)^3$ and therefore 
\be{partialeta}
\partial_A\eta^{\alpha}=\frac{\partial \ln x}{\partial\ln a}\frac{\rd \eta^{\alpha}}{\rd\ln x}=3\frac{\rd \eta^\alpha}{\rd\ln x}=\frac{\partial \ln x}{\partial \ln \phi}\frac{\rd \eta^{\alpha}}{\rd\ln x}=\partial_F\eta^{\alpha}.
\ee
The conditions (\ref{auxdef}) can be then satisfied by setting $\tau^A=-\tau^F=\tau_0^{-1}={\rm const}$. One then has the following auxiliary vector field \be{patau}
\partial_\tau=\tau_0^{-1}\pa{\partial_A-\partial_F}
\ee
which is associated with a new coordinate. The coordinates $(\eta,\tau)$ can now be adopted to parametrise the configuration space and can be related to $(A,F)$ by the following change of variable
\be{chvartau1}
A=\frac{\tau}{\tau_0}+A_{\rm cl}(\eta),\; F=-\frac{\tau}{\tau_0}+F_{\rm cl}(\eta)
\ee
where 
\be{ApF}
A+F=A_{\rm cl}(\eta)+F_{\rm cl}(\eta)
\ee
is a function of $\eta$ only.
By inverting the relations (\ref{wkbtime},\ref{patau}) one has
\be{chvarAF}
\partial_A=\frac{1}{\eta^A+\eta^F}\pa{\partial_\eta+\eta^F\,\tau_0\partial_\tau},\;\partial_F=\frac{1}{\eta^A+\eta^F}\pa{\partial_\eta-\eta^A\,\tau_0\partial_\tau}.
\ee
%Let us note that 
%\be{pataueta}
%\partial_\tau\eta^\alpha=\partial_A\eta^\alpha-\partial_F\eta^\alpha=\pa{\frac{\rd \ln x}{\rd \ln a}-\frac{\rd \ln x}{\rd \ln \phi}}\frac{\partial \eta^\alpha}{\partial\ln x}=\pa{3-3}\frac{\partial \eta^\alpha}{\partial\ln x}=0.
%\ee
Let us note that while $\partial_\eta$ is a vector tangent to the classical trajectories in minisuperspace, $\partial_\tau$ is not associated to any particular direction. 
The two vectors are necessary in order to estimate the quantum gravitational corrections to the MS equation originally parametrised by $(a,\phi)$. Locally, given $\partial_\eta$, one can always find a vector orthogonal to it (here the orthogonality means the orthogonality w.r.t. the minisuperspace supermetric) and the quantum gravitational effects may be ``projected'' on these two directions. Physically $\partial_\eta$ generates the time flow on the classical trajectory and the associated quantum corrections are the fluctuations along such a trajectory. On the other hand the quantum corrections on the orthogonal direction describe the fluctuations away from a given classical trajectory. When one performs the BO decomposition factorising only one homogeneous degree of freedom and then using it as the ``classical clock'' for the rest of the system, by construction only the quantum fluctuations along the classical trajectory are present.\\
If one now considers the de Sitter attractor (\ref{vel4dS}) the above expressions are simplified and one obtains $\eta_F=0$, $\eta_A=\sqrt{\frac{\lambda}{3\xi}}\,a\,\phi $ and
\be{dSchvar}
\partial_A=\frac{1}{\eta^A}\partial_\eta,\;\partial_F=\frac{1}{\eta^A}\partial_\eta-\tau_0\partial_\tau.
\ee
For such a case $\partial_\tau$ given by (\ref{patau}) is orthogonal to $\partial_\eta$ globally. We shall adopt this definition of $\tau$ in the following sections. 
%%%%%%%%%%%%%%%%%%%%%%%%%%%%%%%%
\subsection{The modified MS Equation}
The equation for the wave function of the perturbations is (\ref{keq2}). In such an equation the terms related to the introduction of the time are
\be{timeint}
-\frac{2G^{\alpha\beta}}{a_0^2\re^{2A}h(F)}\frac{\partial_\alpha\widetilde \Psi_0}{\widetilde \Psi_0}\partial_\beta\tilde\chi_k= i\pa{\eta^\alpha\partial_\alpha+q^{\alpha}\partial_\alpha}\tilde\chi_k
\equiv i\pa{\partial_\eta+q^{\alpha}\partial_\alpha}\tilde\chi_k
\ee
and contributions proportional to $i\pa{q^{\alpha}\partial_\alpha}\tilde\chi_k$ should be considered as quantum corrections emerging from the definition of time. Eq. (\ref{keq2}) also contains ``pure'' quantum gravitational contributions (originating from non-adiabatic effects) given by
\be{pureQG}
\pa{\hat Q-\av{\hat{\widetilde{Q}}}}\tilde\chi_k=
\frac{G^{\alpha\beta}}{a_0^2{\rm e}^{2A}h(F)}\paq{\pa{\partial_\alpha\partial_\beta-\av{\widetilde{\partial_\alpha\partial_\beta}}}}\tilde\chi_k.
\ee
Solving the full quantum equation (\ref{keq2}) is a hopeless task. One still may search for a solution perturbatively. To the leading order one has
\be{keq2o0} 
\left\{-i\frac{\partial}{\partial \eta}+\pa{\hat{\mathcal H}_k-\av{\hat{\mathcal H}_k}}\right\}\tilde\chi_k(\eta,v_k)=0
\ee
where, in $\hat {\mathcal H}_k$, $a=a(\eta)$ and $\phi=\phi(\eta)$ are the classical trajectories on minisuperspace. One may now redefine 
\be{defchis}
\chi_{k,s}\equiv \exp\paq{-i\int \rd \eta'\av{\hat{\mathcal H}_k}}\tilde\chi_k
\ee
and obtain the standard MS equation 
\be{keq3o0} 
\pa{i\frac{\partial}{\partial \eta}-\hat{\mathcal H}_k}\chi_{k,s}=0.
\ee
This equation can be solved exactly in some cases (for example on a de Sitter background) or in the Slow Roll approximation. On then following a perturbative approach, the quantum gravitational corrections can be evaluated using the leading order solution. Let us note that the solutions of (\ref{keq3o0}) are functions of $\eta$ and not of the auxiliary parameter $\tau$. Therefore the quantum gravitational effects are only generated by the derivatives with respect to the classical time flow $\partial_\eta$ and not $\partial_\tau$. In the first order equation, all the contributions arising from $\partial_\tau$ can be ignored. If $\partial_\tau$ is chosen as the direction orthogonal to the classical trajectory we conclude that, within the perturbative approach, the fluctuations away from the classical trajectory must be zero.
%%%%%%%%%%%%
\section{Application to de Sitter evolution}
One may apply the method described above to de Sitter evolution and IG. Such a case is relevant as it describes inflation to the leading order in the slow roll approximation and many expressions simplify. One may then easily check how quantum gravitational corrections to the primordial power spectra can be calculated without having too complicated expressions. In IG the stable de Sitter attractor exists only for a quartic potential. Correspondingly the scalar field (at least classically) is constant and takes a value which is arbitrary and only depends on the initial conditions. We shall only consider the solutions corresponding to the above mentioned attractor and ignore those which describe the transient phase with the scalar field slowing down and approaching the attractor asymptotically.\\
Classically the formulae relevant for this case have been presented in the sections 2 and 3. The auxiliary vector has been already calculated and its relation with the coordinate basis vectors on the configuration space are given by (\ref{dSchvar}).\\
The general full perturbation equation is
\be{keq4}
\paq{-i\pa{\eta^{\alpha}+q^{\alpha}}\partial_\alpha+\pa{\hat{\mathcal{H}}_k-\av{\hat{\mathcal{H}}_k}}+\pa{\hat Q-\av{\hat{\widetilde{Q}}}}}\tilde \chi_k=0
\ee
where $\hat Q$ is defined by (\ref{pureQG}) and $q^\alpha$ is defined implicitly by (\ref{timeint}).\\
For the de Sitter attractor, the time derivative is $\partial_\eta=\eta^\alpha\partial_\alpha$ with
\be{dSdef1}
\eta^F=0,\;\eta^A=\sqrt{\frac{\lambda}{3\xi}}\,a\,\phi\equiv a\, H,
\ee 
$H=\sqrt{\frac{\lambda}{3\xi}}\phi$, and the corresponding auxiliary vector is $\partial_\tau$ and they are related to $\partial_A$ and $\partial_F$ by the following relations
\be{dSdef2}
\partial_A=\frac{1}{aH}\partial_\eta,\;\partial_F=\frac{1}{aH}\pa{\partial_\eta-aH\partial_\tau}.
\ee
Let us note that $a H=\sqrt{\frac{\lambda}{3\xi}}\exp\pa{A+F}$ and, see (\ref{ApF}), is a function of $\eta$ only. 
Moreover when $\partial_F$ acts on a function of $\eta$ (and not $\tau$) one has $\partial_F f(\eta)= \frac{1}{aH}\partial_\eta f(\eta)$.
Therefore
\be{gabdS}
G^{\alpha\beta}\partial_\alpha\partial_\beta=\frac{1+6\xi}{12\xi}\pa{\frac{1}{a^2H^2}\partial_\eta^2-\frac{1}{aH}\partial_\eta}
\ee
and the perturbations equation then becomes
\ba{keq5}
&&\pag{-i \pa{1+\frac{q^A+q^F}{aH}}\partial_\eta+\pa{\hat{\mathcal{H}}_k-\av{\hat{\mathcal{H}}_k}}\right.\nonumber\\
&&\left.+\frac{1}{12\xi a^4H^2\phi^2}\paq{\partial_\eta^2-\av{\widetilde{\partial_\eta^2}}-aH\pa{\partial_\eta-\av{\widetilde{\partial_\eta}}}}\tilde\chi_k
},
\ea
where the quantities $a\,H$ and $a\, \phi$ are functions of $\eta$ evaluated on the inflationary attractor. Let us note that, to the leading order, the above equation becomes (\ref{keq2o0}) and $\av{\widetilde{\partial_\eta}}=0$.\\
Let us now evaluate the quantum corrections associated with the introduction of time. They are given by 
\be{qctime}
\pa{\hat T-\av{\hat{\widetilde{T}}}}\tilde\chi_k\equiv -i \pa{\frac{q^A+q^F}{aH}}\pa{\partial_\eta-\av{\widetilde\partial_\eta}}\tilde\chi_k,
\ee
 where the quantities $q^\alpha$ are implicitly defined by (\ref{timeint}) and are thus given by
 \be{defq}
 q^\beta=\frac{2i}{a^2h}G^{\alpha\beta}\frac{\partial_\alpha\tilde\Psi_0}{\tilde\Psi_0}-\eta^\beta.
 \ee
On the inflationary attractor $\tilde\Psi_0$ is a function of $x\equiv a^3\phi^3$ (see \cite{IGwdw} for the details). Let us rephase it as
\be{psi0rep}
\tilde\Psi_0\equiv \psi_q\exp\pa{i W},
\ee
where $W$ is the Hamilton Jacobi function which satisfies (\ref{HJeq}) with $\nu=0$ and $\tilde\Psi_0$ satisfies the homogeneous WdW (\ref{homeq2}) where, for simplicity, we neglect the backreaction of the perturbations (recovering the WdW equation in minisuperspace). The rephased wave function $\psi_q$ satisfies the following equation
\be{Psiqeq}
2i\frac{\rd W}{\rd \ln x}\frac{\rd \ln \psi_q}{\rd \ln x}+i\frac{\rd^2 W}{\rd \ln x^2}+\frac{\rd^2\ln \psi_q}{\rd \ln x^2}+\pa{\frac{\rd \ln \psi_q}{\rd \ln x}}^2=0
\ee
where the first two terms usually give the leading contribution in the semiclassical ($\hbar\rightarrow 0$) expansion and lead to the van Vleck determinant. One finally obtains
\be{qexpressions}
q^A=\frac{i}{2\xi a^2\phi^2}\frac{\rd \ln \psi_q}{\rd \ln x}, \;q^F=0
\ee
for the inflationary attractor we are considering.
In contrast, on neglecting the last two terms in (\ref{Psiqeq}) (and then following the prescription for the standard WKB approximation), the expression for $\psi_q$ can be easily calculated in terms of $W$ and one has $\rd \ln \psi_q/\rd \ln x= -1/2$. In this latter case, the quantum effects are inversely proportional to $\xi\phi^2$. Let us note that $q^A$ only depends on $x$ and is therefore a function of $\eta$ only. \\
The existence of exact solutions for IG and a power law potential also allows an explicit calculation of the above quantum gravitational corrections. In \cite{IGwdw} we found the following exact solution for IG with a quartic potential:
\be{exactsol}
\tilde \Psi_0=\pa{\frac{a}{a_0}}^\nu \chi(x),
\ee
where
\be{chiex}
\chi(x)=x^q\paq{c_1J_r(Ax)+c_2 Y_r(A x)}
\ee
with $x=a^3\phi^3$, $A=\pa{\frac{4}{3}\xi\lambda}^{1/2}$, $r=q=0$, and $J_r$, $Y_r$ are Bessel functions. Let us note that the superpositions of the Bessel functions generally mix contracting and expanding universes. The solution corresponding to the classical evolution on the de Sitter attractor corresponds to $\nu=0$. In the limit for large $\z\equiv A x=2\xi\phi^2 a^3H$ one has
\be{asyJ}
J_r(\z)\sim \frac{1}{\sqrt{2\pi \z}}\paq{{\rm e}^{i\pa{\z-\frac{\pi}{4}}}\pa{1-\frac{i}{8\z}-\frac{9}{128\z^2}+\mathcal{O}\pa{\frac{1}{\z^3}}}+\rm{c.c.}}
\ee
and
\be{asyJ}
Y_r(\z)\sim \frac{1}{\sqrt{2\pi \z}}\paq{-i{\rm e}^{i\pa{\z-\frac{\pi}{4}}}\pa{1-\frac{i}{8\z}-\frac{9}{128\z^2}+\mathcal{O}\pa{\frac{1}{\z^3}}}+\rm{c.c.}}.
\ee
Let us now consider the linear combination with $c_2=-i c_1$ which corresponds to the expanding phase.
Then, following the procedure for the introduction of time, we find
\be{emtimeds1}
-\frac{2G^{\alpha\beta}}{a^2h}\frac{\partial_\alpha\tilde\Psi_0}{\tilde\Psi_0}\partial_\beta=-\frac{\z}{2\xi a\phi^2}\frac{\partial_\z\chi}{\chi}\partial_a.
\ee
\\
If we keep the leading and next to leading contribution in 
\be{pachi}
\frac{\partial_\z\chi}{\chi}\sim -i\pa{1-\frac{i}{2\z}+\dots}
\ee
for large $\z$ then
\be{emtimeds2}
-\frac{\z}{2\xi a\phi^2}\frac{\partial_\z\chi}{\chi}\partial_a\simeq ia^2H\pa{1-\frac{i}{2\z}}\partial_a\simeq i\partial_\eta+\frac{1}{2\z}\partial_\eta
\ee
and thus
\be{identq}
iq^\alpha\partial_\alpha=\frac{1}{2\z}\partial_\eta.
\ee
This last contribution is proportional to $\pa{\xi\phi^2}$ and is identical to that obtained with the WKB approximation.\\ 
Let us note that the quantum gravitational corrections evaluated in terms of $\eta$ and the auxiliary variable $\tau$ only depend on $a\phi$, which is a function of $\eta$ only, and on the first and second order derivatives in $\partial_\eta$ and $\partial_\tau$. Therefore the resulting ``modified'' MS equation admits solutions of the form $\chi_{k,s}=\chi_{k,s}(\eta)$ (without any functional dependence on $\tau$). Therefore, to the leading order, one recovers the usual MS equation having solutions which depend on the classical time $\eta$. To the next to leading order the quantum gravitational corrections are evaluated perturbatively with the leading order solution and therefore the perturbed solution is a function of $\eta$ only. Thus the quantum gravitational corrections associated to the direction orthogonal to the time flow are necessarily zero and one is left with those parallel to the classical trajectory.\\
Finally one can rephase $\tilde\chi_k$ according to the prescription (\ref{defchis}), express (\ref{keq5}) in terms of $\chi_{k,s}$ (which satisfies the conventional Schr\"odinger equation (\ref{keq3o0}) to leading order) and obtain 
\ba{keq6}
&&\left\{\pa{-i\frac{\partial}{\partial \eta}+\hat{\mathcal{H}_k}}-\frac{i}{2\xi a^3H\phi^2}\frac{\rd \ln \psi_q}{\rd \ln x}\pa{\hat{\mathcal{H}_k}-\av{\hat{\mathcal{H}_k}}_s}+\right.\nonumber\\
&&\left.\frac{1}{12\xi a^4H^2\phi^2}\pa{\av{i\partial_\eta \hat{\mathcal{H}_k}}_s-i\partial_\eta \hat{\mathcal{H}_k}}-\pa{\hat{\mathcal{H}_k}-\av{\hat{\mathcal{H}_k}}_s}^2+\right.\nonumber\\
&&\left.\av{\hat{\mathcal{H}^2}}_s-\av{\hat{\mathcal{H}_k}}_s^2+i aH\pa{\hat{\mathcal{H}_k}-\av{\hat{\mathcal{H}_k}}_s}
\right\}\chi_{k,s}=0,
\ea
where now $\av{\hat O}_s\equiv \av{\chi_{k,s}|\hat O|\chi_{k,s}}$.
On also considering the van Vleck contribution to the introduction of time and defining $\xi\phi^2=\tilde m_{\rm P}^2/6$ (the effective value of the Planck mass in the IG framework) one obtains
\be{keqDS}
\pa{i\frac{\partial }{\partial \eta}-\hat{\mathcal{H}_k}}\chi_s=\frac{1}{2\tilde m_{\rm P}^2}\pa{\hat \Omega_k-\av{\hat \Omega_k}_s}\chi_s,
\ee
where 
\be{defOmegak}
\hat \Omega_k=\frac{1}{a^4H^2}\paq{2\av{\hat{\mathcal{H}_k}}_s\hat{\mathcal{H}_k}-\hat{\mathcal{H}_k}^2-i\frac{\rd \hat{\mathcal{H}_k}}{\rd \eta}+4\pa{a H} \hat{\mathcal{H}_k}}.
\ee
Let us note that formally this result is the same as the one for the de Sitter solution in GR with the identification of $\widetilde{m}_{\rm P}$ and $m_{\rm P}\equiv \sqrt{6}\M$ where the former is proportional to the effective Planck mass, which depends on the expectation value of the scalar (inflaton) field and the latter is proportional to the Planck mass.\\
Furthermore on expressing the modified MS equation (\ref{keqDS}) in terms of the Einstein Frame DOF $\tilde a$, $\tilde \phi$, $\tilde\eta$, $\tilde v_k$ with
\be{EFdof}
\tilde a=\frac{\sqrt{6\xi}}{m_{\rm P}a\,\phi},\;\tilde \phi=\sqrt{\frac{1+6\xi}{6\xi}}m_{\rm P}\ln \frac{\phi}{\M},\;\tilde H=\frac{m_{\rm P}}{\sqrt{6\xi}}\frac{H}{\phi}
\ee
we observe that
\be{EFrel}
a\,H=\tilde a\tilde H\Rightarrow \eta=\tilde \eta\quad {\rm and}\quad v_k=\tilde v_k
\ee
and therefore one recovers exactly the equation already found for GR \cite{BOarts}.
We can therefore conclude that, on even including the quantum gravitational corrections and in the ``pure'' de Sitter case, the primordial spectra are invariant w.r.t. the Jordan to Einstein frame transformation. Indeed the de Sitter evolution is invariant with respect to frame transformations and the primordial spectra calculated without the quantum gravitational corrections are the same (this latter property of the primordial spectra is valid independently of the background evolution chosen). Such an invariance holds also when quantum gravitational corrections are included. Let us note that the fact that such an equivalence holds at the quantum level (at least for the de Sitter case) also confirms the consistency of the approach adopted here for the introduction of time in a matter-gravity system with two minisuperspace variables playing the role of the ``classical clock''.\\
%%%%%%%%%%%%
\section{Conclusions}
Non-minimally coupled scalar fields are ubiquitous in cosmology, in particular when energies become very high since a non-minimal coupling generally emerges from quantum effects. It seems therefore natural to study their quantum behaviour (in particular during inflation with the scalar field playing the role of the inflaton) in the presence of the quantum gravitational effects which are usually ignored (or included in an effective description) in the inflationary era and calculate the evolution of the inflationary spectra. Theories with non-minimally coupled scalar fields are usually included in the class of modified gravity theories since such scalar fields affect Newton's constant and can modify gravitational attraction even at long distances. Furthermore there exists a mapping between the DOF of such theories and those of General Relativity with a minimally coupled scalar field, which is called Jordan to Einstein frame mapping. The mapping is often used since performing calculations in the Einstein Frame is usually easier and the results can be finally translated into the Jordan frame through the inverse mapping. This ``equivalence'' is known to hold at classical and semiclassical levels but at the full quantum level the complete equivalence of the two frames is not clear \cite{JFEF}.\\
In this article the technique already employed in a series of articles \cite{BO} for a minimally coupled inflaton and standard General Relativity is applied to inflation with a non-minimally coupled inflaton. Such a technique leads to a MS equation with quantum gravitational corrections. The resulting quantum corrections can then be calculated explicitly for different inflationary models and the resulting inflationary spectra obtained. Moreover the full quantum equivalence between the Einstein and the Jordan frame can be investigated case by case (at least in the canonical quantisation context and within the approximation scheme followed). As an application we calculated the corrections on a de Sitter background. The MS equation obtained reproduces correctly that of \cite{BO} in the minimally coupled limit and the resulting spectra are invariant in both frames. This latter result is a consequence of the fact that the de Sitter evolution is frame invariant and is non-trivial.\\ 
Furthermore we discussed the problem of the introduction of time in the context of quantum cosmology by generalising the approach adopted in \cite{BO}. The full scheme presented can be applied to cases more general than pure de Sitter, in particular more general inflationary potentials should be considered and more realistic inflationary evolutions (including first order correction in the slow roll approximation) studied as was already done for the minimally coupled case \cite{BOarts}.

%%%%%%%%%%%%%%%%%%%%%%%
\section{Acknowledgements}
Alexander Y. Kamenshchik is supported in part by the RFBR grant 18-52-45016.

%%%%%%%%%%%%%%%%%%

%%%%%%%%%%%%%%%%%%%%

\begin{thebibliography}{99}

\bibitem{inflation}
A.A. Starobinsky. Springer. in H.J. De Vega and N. Sanchez (eds.) Current trends in field theory quantum gravity and strings, Lecture Notes in Physics 246 Verlag, Heidelberg, 1986), pp. 107-126.
	A.D. Linde. Academic. Particle Physics and Inflationary Cosmology (Harwood New York, 1990).
	
 \bibitem{Planck}
 P.~A.~R.~Ade {\it et al.} [Planck Collaboration],
 %``Planck 2015 results. XX. Constraints on inflation,''
 arXiv:1502.02114 [astro-ph.CO].
 %%CITATION = ARXIV:1502.02114;%%
 %682 citations counted in INSPIRE as of 15 Jun 2016
 P.~A.~R.~Ade {\it et al.} [Planck Collaboration],
 %``Planck 2015 results. XIII. Cosmological parameters,''
 arXiv:1502.01589 [astro-ph.CO].
 %%CITATION = ARXIV:1502.01589;%%
 Y.~Akrami {\it et al.} [Planck Collaboration],
 %``Planck 2018 results. X. Constraints on inflation,''
 arXiv:1807.06211 [astro-ph.CO].
 %%CITATION = ARXIV:1807.06211;%%
 %112 citations counted in INSPIRE as of 09 Nov 2018
N.~Aghanim {\it et al.} [Planck Collaboration],
 %``Planck 2018 results. VI. Cosmological parameters,''
 arXiv:1807.06209 [astro-ph.CO].
 %%CITATION = ARXIV:1807.06209;%%
 %239 citations counted in INSPIRE as of 09 Nov 2018
 
 \bibitem{pert}
V.F. Mukhanov, Sov. Phys. JETP {\bf 68}, 1297 (1988);
J.~M.~Maldacena,
 %``Non-Gaussian features of primordial fluctuations in single field inflationary models,''
 JHEP {\bf 0305} (2003) 013;
 V.~F.~Mukhanov, H.~A.~Feldman and R.~H.~Brandenberger,
 %``Theory of cosmological perturbations. Part 1. Classical perturbations. Part 2. Quantum theory of perturbations. Part 3. Extensions,''
 Phys.\ Rept.\ {\bf 215} (1992) 203.
 V.F. Mukhanov, Phys. Lett. B {\bf 218}, 17 (1989);
 J.~M.~Bardeen,
 %``Gauge Invariant Cosmological Perturbations,''
 Phys.\ Rev.\ D {\bf 22}, 1882 (1980).
 doi:10.1103/PhysRevD.22.1882
 M.~Sasaki,
  %``Gauge Invariant Scalar Perturbations in the New Inflationary Universe,''
  Prog.\ Theor.\ Phys.\  {\bf 70} (1983) 394.
  doi:10.1143/PTP.70.394
  %%CITATION = doi:10.1143/PTP.70.394;%%
  %63 citations counted in INSPIRE as of 21 Nov 2019
 
 \bibitem{BHform}
B.~J.~Carr,
 %``The Primordial black hole mass spectrum,''
 Astrophys.\ J.\ {\bf 201} (1975) 1.
 doi:10.1086/153853
 %%CITATION = doi:10.1086/153853;%%
 %503 citations counted in INSPIRE as of 09 Nov 2018
 M.~Sasaki, T.~Suyama, T.~Tanaka and S.~Yokoyama,
 %``Primordial black holes?perspectives in gravitational wave astronomy,''
 Class.\ Quant.\ Grav.\ {\bf 35} (2018) no.6, 063001
 doi:10.1088/1361-6382/aaa7b4
 [arXiv:1801.05235 [astro-ph.CO]].
 %%CITATION = doi:10.1088/1361-6382/aaa7b4;%%
 %41 citations counted in INSPIRE as of 09 Nov 2018
 A.~Y.~Kamenshchik, A.~Tronconi, T.~Vardanyan and G.~Venturi,
 %``Non-Canonical Inflation and Primordial Black Holes Production,''
 Phys.\ Lett.\ B {\bf 791}, 201 (2019)
 doi:10.1016/j.physletb.2019.02.036
 [arXiv:1812.02547 [gr-qc]].
 %%CITATION = doi:10.1016/j.physletb.2019.02.036;%%
 %4 citations counted in INSPIRE as of 28 Aug 2019
 
 \bibitem{AS}
 K.~G.~Wilson,
 %``Renormalization group and critical phenomena. 1. Renormalization group and the Kadanoff scaling picture,''
 Phys.\ Rev.\ B {\bf 4} (1971) 3174.
 doi:10.1103/PhysRevB.4.3174
 %%CITATION = doi:10.1103/PhysRevB.4.3174;%%
 %634 citations counted in INSPIRE as of 09 Nov 2018
 K.~G.~Wilson,
 %``Renormalization group and critical phenomena. 2. Phase space cell analysis of critical behavior,''
 Phys.\ Rev.\ B {\bf 4} (1971) 3184.
 doi:10.1103/PhysRevB.4.3184
 %%CITATION = doi:10.1103/PhysRevB.4.3184;%%
 %541 citations counted in INSPIRE as of 09 Nov 2018
S.~Weinberg,
 %``Effective Field Theory, Past and Future,''
 PoS CD {\bf 09}, 001 (2009)
 [arXiv:0908.1964 [hep-th]].
 %%CITATION = ARXIV:0908.1964;%%
 %103 citations counted in INSPIRE as of 16 Mar 2017
M.~Reuter,
 %``Nonperturbative evolution equation for quantum gravity,''
 Phys.\ Rev.\ D {\bf 57}, 971 (1998)
 [hep-th/9605030].
 %%CITATION = doi:10.1103/PhysRevD.57.971;%%
 %543 citations counted in INSPIRE as of 16 Mar 2017
 A.~Tronconi,
 %``Asymptotically Safe Non-Minimal Inflation,''
 JCAP {\bf 1707} (2017) no.07, 015
 doi:10.1088/1475-7516/2017/07/015
 [arXiv:1704.05312 [gr-qc]].
 %%CITATION = doi:10.1088/1475-7516/2017/07/015;%%
 %8 citations counted in INSPIRE as of 28 Aug 2019
 A.~Bonanno, A.~Platania and F.~Saueressig,
 %``Cosmological bounds on the field content of asymptotically safe gravity?matter models,''
 Phys.\ Lett.\ B {\bf 784} (2018) 229
 doi:10.1016/j.physletb.2018.06.047
 [arXiv:1803.02355 [gr-qc]].
 %%CITATION = doi:10.1016/j.physletb.2018.06.047;%%
 %10 citations counted in INSPIRE as of 12 Apr 2019
 A.~Eichhorn,
 %``Status of the asymptotic safety paradigm for quantum gravity and matter,''
 Found.\ Phys.\ {\bf 48} (2018) no.10, 1407
 doi:10.1007/s10701-018-0196-6
 [arXiv:1709.03696 [gr-qc]].
 %%CITATION = doi:10.1007/s10701-018-0196-6;%%
 %42 citations counted in INSPIRE as of 12 Apr 2019
 
 \bibitem{higgsinf}
 F.~L.~Bezrukov and M.~Shaposhnikov,
 %``The Standard Model Higgs boson as the inflaton,''
 Phys.\ Lett.\ B {\bf 659} (2008) 703
 doi:10.1016/j.physletb.2007.11.072
 [arXiv:0710.3755 [hep-th]].
 %%CITATION = doi:10.1016/j.physletb.2007.11.072;%%
 %1214 citations counted in INSPIRE as of 12 Apr 2019
 F.~L.~Bezrukov, A.~Magnin and M.~Shaposhnikov,
  %``Standard Model Higgs boson mass from inflation,''
  Phys.\ Lett.\ B {\bf 675} (2009) 88
  doi:10.1016/j.physletb.2009.03.035
  [arXiv:0812.4950 [hep-ph]].
  %%CITATION = doi:10.1016/j.physletb.2009.03.035;%%
  %267 citations counted in INSPIRE as of 25 Nov 2019
 A.~O.~Barvinsky, A.~Y.~Kamenshchik and A.~A.~Starobinsky,
  %``Inflation in the Standard Model with a strong non-minimal curvature coupling and the Higgs boson mass,''
  %%CITATION = INSPIRE-1372906;%%
  A.~O.~Barvinsky, A.~Y.~Kamenshchik, C.~Kiefer, A.~A.~Starobinsky and C.~Steinwachs,
  %``Asymptotic freedom in inflationary cosmology with a non-minimally coupled Higgs field,''
  JCAP {\bf 0912} (2009) 003
  doi:10.1088/1475-7516/2009/12/003
  [arXiv:0904.1698 [hep-ph]].
  %%CITATION = doi:10.1088/1475-7516/2009/12/003;%%
  %215 citations counted in INSPIRE as of 25 Nov 2019
  A.~De Simone, M.~P.~Hertzberg and F.~Wilczek,
  %``Running Inflation in the Standard Model,''
  Phys.\ Lett.\ B {\bf 678} (2009) 1
  doi:10.1016/j.physletb.2009.05.054
  [arXiv:0812.4946 [hep-ph]].
  %%CITATION = doi:10.1016/j.physletb.2009.05.054;%%
  %337 citations counted in INSPIRE as of 25 Nov 2019
 
 \bibitem{induced}
A. D. Sakharov, Dokl. Akad. Nauk. SSSR 117, 70 (1967); 
[Sov. Phys. Dokl. 12, 1040 (1967)];
 A.~Zee,
 %``Spontaneously Generated Gravity,''
 Phys.\ Rev.\ D {\bf 23}, 858 (1981);
 %%CITATION = PHRVA,D23,858;%%
 
 \bibitem{cooper}
 F.~Cooper and G.~Venturi,
 %``Cosmology and Broken Scale Invariance,''
 Phys.\ Rev.\ D {\bf 24} (1981) 3338.
 doi:10.1103/PhysRevD.24.3338
 %%CITATION = doi:10.1103/PhysRevD.24.3338;%%
 %54 citations counted in INSPIRE as of 12 Apr 2019
 F.~Finelli, A.~Tronconi and G.~Venturi,
 %``Dark Energy, Induced Gravity and Broken Scale Invariance,''
 Phys.\ Lett.\ B {\bf 659} (2008) 466
 doi:10.1016/j.physletb.2007.11.053
 [arXiv:0710.2741 [astro-ph]].
 %%CITATION = doi:10.1016/j.physletb.2007.11.053;%%
 %34 citations counted in INSPIRE as of 12 Apr 2019
 A.~Tronconi and G.~Venturi,
 %``Quantum Back-Reaction in Scale Invariant Induced Gravity Inflation,''
 Phys.\ Rev.\ D {\bf 84} (2011) 063517
 doi:10.1103/PhysRevD.84.063517
 [arXiv:1011.3958 [gr-qc]].
 %%CITATION = doi:10.1103/PhysRevD.84.063517;%%
 %17 citations counted in INSPIRE as of 12 Apr 2019
 A.~Y.~Kamenshchik, A.~Tronconi and G.~Venturi,
 %``Dynamical Dark Energy and Spontaneously Generated Gravity,''
 Phys.\ Lett.\ B {\bf 713} (2012) 358
 doi:10.1016/j.physletb.2012.06.035
 [arXiv:1204.2625 [gr-qc]].
 %%CITATION = doi:10.1016/j.physletb.2012.06.035;%%
 %18 citations counted in INSPIRE as of 12 Apr 2019
 A.~Cerioni, F.~Finelli, A.~Tronconi and G.~Venturi,
 %``Inflation and Reheating in Induced Gravity,''
 Phys.\ Lett.\ B {\bf 681} (2009) 383
 doi:10.1016/j.physletb.2009.10.066
 [arXiv:0906.1902 [astro-ph.CO]];
 A.~Cerioni, F.~Finelli, A.~Tronconi and G.~Venturi,
 %``Inflation and Reheating in Spontaneously Generated Gravity,''
 Phys.\ Rev.\ D {\bf 81} (2010) 123505
 doi:10.1103/PhysRevD.81.123505
 [arXiv:1005.0935 [gr-qc]].
 %%CITATION = doi:10.1103/PhysRevD.81.123505;%%
 %24 citations counted in INSPIRE as of 12 Apr 2019
 %%CITATION = doi:10.1016/j.physletb.2009.10.066;%%
 %43 citations counted in INSPIRE as of 12 Apr 2019
 
 \bibitem{BOarts}
A.~Y.~Kamenshchik, A.~Tronconi and G.~Venturi,
 %``Inflation and Quantum Gravity in a Born-Oppenheimer Context,''
 Phys.\ Lett.\ B {\bf 726} (2013) 518
 doi:10.1016/j.physletb.2013.08.067
 [arXiv:1305.6138 [gr-qc]].
 %%CITATION = doi:10.1016/j.physletb.2013.08.067;%%
 %26 citations counted in INSPIRE as of 28 Aug 2019
 A.~Y.~Kamenshchik, A.~Tronconi and G.~Venturi,
 %``Signatures of quantum gravity in a Born?Oppenheimer context,''
 Phys.\ Lett.\ B {\bf 734} (2014) 72
 doi:10.1016/j.physletb.2014.05.028
 [arXiv:1403.2961 [gr-qc]].
 %%CITATION = doi:10.1016/j.physletb.2014.05.028;%%
 %23 citations counted in INSPIRE as of 28 Aug 2019
 A.~Y.~Kamenshchik, A.~Tronconi and G.~Venturi,
 %``Quantum Gravity and the Large Scale Anomaly,''
 JCAP {\bf 1504} (2015) no.04, 046
 doi:10.1088/1475-7516/2015/04/046
 [arXiv:1501.06404 [gr-qc]].
 %%CITATION = doi:10.1088/1475-7516/2015/04/046;%%
 %13 citations counted in INSPIRE as of 28 Aug 2019
 A.~Y.~Kamenshchik, A.~Tronconi and G.~Venturi,
 %``Quantum Cosmology and the Evolution of Inflationary Spectra,''
 Phys.\ Rev.\ D {\bf 94} (2016) no.12, 123524
 doi:10.1103/PhysRevD.94.123524
 [arXiv:1609.02830 [gr-qc]].
 %%CITATION = doi:10.1103/PhysRevD.94.123524;%%
 %10 citations counted in INSPIRE as of 28 Aug 2019
A.~Y.~Kamenshchik, A.~Tronconi and G.~Venturi,
 %``The Born?Oppenheimer method, quantum gravity and matter,''
 Class.\ Quant.\ Grav.\ {\bf 35} (2018) no.1, 015012
 doi:10.1088/1361-6382/aa8fb3
 [arXiv:1709.10361 [gr-qc]].
 %%CITATION = doi:10.1088/1361-6382/aa8fb3;%%
 %11 citations counted in INSPIRE as of 28 Aug 2019
 
 \bibitem{DeWitt}
B.S. DeWitt, Phys. Rev. {\bf 160}, 113 (1967)

\bibitem{BO}
M. Born and J.R. Oppenheimer, Ann. Physik {\bf 84}, 457 (1927); 
C. A. Mead and D. G. Truhlar, J. Chem. Phys. {\bf 70}, 2284 (1979);
C. A. Mead, Chem. Phys {\bf 49}, 23 (1980)
C. A. Mead, Chem. Phys {\bf 49}, 33 (1980)
R. Brout and G. Venturi, Phys. Rev. D {\bf 39}, 2436 (1989);
G. Venturi, Class. Quantum Grav. {\bf 7}, 1075 (1990);
G.~L.~Alberghi, R.~Casadio, A.~Tronconi,
 %``Planck scale inflationary spectra from quantum gravity,''
 Phys.\ Rev.\ D {\bf 74} (2006) 103501;
G.~L.~Alberghi, C.~Appignani, R.~Casadio, F.~Sbisa, A.~Tronconi,
 %``Inflation and the semiclassical dynamics of a conformal scalar field,''
 Phys.\ Rev.\ D {\bf 77} (2008) 044002;
 A.~Tronconi, G.~P.~Vacca and G.~Venturi,
 %``The Inflaton and time in the matter gravity system,''
 Phys.\ Rev.\ D {\bf 67}, 063517 (2003)
 A.~Tronconi, G.~P.~Vacca and G.~Venturi,
 %``The Inflaton and time in the matter gravity system,''
 Phys.\ Rev.\ D {\bf 67} (2003) 063517
 doi:10.1103/PhysRevD.67.063517
 [gr-qc/0302030].
 %%CITATION = doi:10.1103/PhysRevD.67.063517;%%
 %16 citations counted in INSPIRE as of 28 Aug 2019
A.~Y.~Kamenshchik, A.~Tronconi, T.~Vardanyan and G.~Venturi,
 %``Quantum Gravity, Time, Bounces and Matter,''
 Phys.\ Rev.\ D {\bf 97} (2018) no.12, 123517
 doi:10.1103/PhysRevD.97.123517
 [arXiv:1804.10075 [gr-qc]].
 %%CITATION = doi:10.1103/PhysRevD.97.123517;%%
 %7 citations counted in INSPIRE as of 28 Aug 2019
 A.~Y.~Kamenshchik, A.~Tronconi, T.~Vardanyan and G.~Venturi,
  %``Time in quantum theory, the Wheeler?DeWitt equation and the Born?Oppenheimer approximation,''
  Int.\ J.\ Mod.\ Phys.\ D {\bf 28} (2019) no.06,  1950073
  doi:10.1142/S0218271819500731
  [arXiv:1809.08083 [gr-qc]].
  %%CITATION = doi:10.1142/S0218271819500731;%%
  %6 citations counted in INSPIRE as of 25 Nov 2019

\bibitem{IGwdw}
A.~Y.~Kamenshchik, A.~Tronconi and G.~Venturi,
 %``Induced Gravity and Quantum Cosmology,''
 Phys.\ Rev.\ D {\bf 100} (2019) no.2, 023521
 doi:10.1103/PhysRevD.100.023521
 [arXiv:1905.02454 [gr-qc]].
 %%CITATION = doi:10.1103/PhysRevD.100.023521;%%

\bibitem{JFEF}
A.~Y.~Kamenshchik, E.~O.~Pozdeeva, S.~Y.~Vernov, A.~Tronconi and G.~Venturi,
  %``Transformations between Jordan and Einstein frames: Bounces, antigravity, and crossing singularities,''
  Phys.\ Rev.\ D {\bf 94} (2016) no.6,  063510
  doi:10.1103/PhysRevD.94.063510
  [arXiv:1602.07192 [gr-qc]].
  %%CITATION = doi:10.1103/PhysRevD.94.063510;%%
  %24 citations counted in INSPIRE as of 21 Nov 2019
  A.~Y.~Kamenshchik, E.~O.~Pozdeeva, A.~Tronconi, G.~Venturi and S.~Y.~Vernov,
  %``Integrable cosmological models with non-minimally coupled scalar fields,''
  Class.\ Quant.\ Grav.\  {\bf 31} (2014) 105003
  doi:10.1088/0264-9381/31/10/105003
  [arXiv:1312.3540 [hep-th]].
  %%CITATION = doi:10.1088/0264-9381/31/10/105003;%%
  %43 citations counted in INSPIRE as of 21 Nov 2019
  A.~Y.~Kamenshchik, E.~O.~Pozdeeva, A.~Tronconi, G.~Venturi and S.~Y.~Vernov,
  %``Interdependence between integrable cosmological models with minimal and non-minimal coupling,''
  Class.\ Quant.\ Grav.\  {\bf 33} (2016) no.1,  015004
  doi:10.1088/0264-9381/33/1/015004
  [arXiv:1509.00590 [gr-qc]].
  %%CITATION = doi:10.1088/0264-9381/33/1/015004;%%
  %23 citations counted in INSPIRE as of 21 Nov 2019
  A.~Y.~Kamenshchik and C.~F.~Steinwachs,
  %``Question of quantum equivalence between Jordan frame and Einstein frame,''
  Phys.\ Rev.\ D {\bf 91} (2015) no.8,  084033
  doi:10.1103/PhysRevD.91.084033
  [arXiv:1408.5769 [gr-qc]].
  %%CITATION = doi:10.1103/PhysRevD.91.084033;%%
  %96 citations counted in INSPIRE as of 21 Nov 2019
 
 
 
\end{thebibliography}
\end{document}